\documentclass[reprint,aps,prd,superscriptaddress,preprintnumbers,nofootinbib]{revtex4-1}

\usepackage{mathrsfs} 
\usepackage{amsfonts}
\usepackage{amsmath}
\usepackage{amsthm}
\usepackage{amssymb}
\usepackage{array}
\usepackage{dcolumn}
\usepackage[normalem]{ulem}
\usepackage[svgnames]{xcolor}
\usepackage[hidelinks]{hyperref}
\hypersetup{
    colorlinks=true,
    linkcolor=NavyBlue,
    filecolor=NavyBlue,
    urlcolor=NavyBlue,
    citecolor=NavyBlue,
    }

\usepackage{graphics}
\usepackage{graphicx}
\usepackage[caption=false]{subfig}
\usepackage{adjustbox}
\usepackage{breqn}
\usepackage{fontawesome} 
\usepackage{slashed}
\usepackage{orcidlink}
\usepackage{afterpage}

\newcommand\lsim{\lesssim}

\def\be{\begin{equation}}
\def\ee{\end{equation}}
\def\bea{\begin{eqnarray}}
\def\eea{\end{eqnarray}}

\newcount\Comments  
\newcommand{\kibitz}[2]{\ifnum\Comments=1\textcolor{#1}{#2}\fi}

\makeatletter
\let\cat@comma@active\@empty
\makeatother

\begin{document}

\title{Supernova Gamma-Ray Constraints from Heavy Sterile Neutrino Decays}

\author{Garv Chauhan\,\orcidlink{0000-0002-8129-8034}}
 \affiliation{Department of Physics, Arizona State University, 450 E. Tyler Mall, Tempe, AZ 85287-1504 USA}
\affiliation{Center for Neutrino Physics, Department of Physics, Virginia Tech, Blacksburg, Virginia 24601, USA}

\author{R. Andrew Gustafson\,\orcidlink{
0000-0002-4794-7459}}
\affiliation{Center for Neutrino Physics, Department of Physics, Virginia Tech, Blacksburg, Virginia 24601, USA}
\affiliation{International Center for Quantum-field Measurement Systems for Studies of the Universe and Particles (QUP,WPI), High Energy Accelerator Research Organization (KEK), Oho 1-1, Tsukuba, Ibaraki 305-081, Japan}

\author{Ian M.~Shoemaker\,\orcidlink{
0000-0001-5434-3744}}
\affiliation{Center for Neutrino Physics, Department of Physics, Virginia Tech, Blacksburg, Virginia 24601, USA}

\begin{abstract} 
Heavy sterile neutrinos can be produced in core-collapse supernovae (CCSNe), which are superb particle generators because of their high densities and temperatures. If the sterile neutrinos are long-lived, these may be produced inside the supernova core and escape the stellar envelope, later decaying into SM particles like photons and neutrinos. In this work, we first improve the calculation of the $\gamma$-ray fluxes. We then revisit the bounds on the sterile neutrino parameter space from the non-observation of $\gamma$-rays from SN1987A by the Solar Maximum Mission (SMM) and constraints from the diffuse $\gamma$-ray background arising from sterile neutrino decays. We find that the constraints arising from both the SMM data and the diffuse $\gamma$-ray background are weaker than those that have previously appeared in the literature. Finally, we study the sensitivity of several present and near-future $\gamma$-ray telescopes such as e-ASTROGAM and Fermi-LAT, assuming a nearby future galactic CCSN. We show that future observations can probe mixing angles as low as $|U_{\tau/\mu4}|^2\sim 5\times10^{-17}$.
\end{abstract}

\maketitle

\section{Introduction}
\noindent
The observation of neutrino signals from SN1987A has proven to be one of the most important astrophysical probes for new physics. A core-collapse supernova is one of the strongest sources of neutrinos, therefore making it especially ideal to test beyond the Standard Model neutrino interactions. One of the most studied possibilities is to add new ``sterile'' neutrinos that mass-mix with the three SM active neutrinos, but do not directly couple to SM gauge bosons. The heaviness of these sterile neutrinos can explain the smallness of the observed neutrino masses via the \textit{seesaw} mechanism~\cite{Minkowski:1977sc, Gell-Mann:1979vob, Mohapatra:1979ia}. While many models of neutrino masses exist, this seesaw scenario provides a well-motivated theoretical target for the on-going experimental effort to discover sterile neutrinos. Significantly, these new states are not entirely sterile since they inherit a mixing suppressed coupling to the electroweak gauge bosons. As a result, they can be searched for in a variety of terrestrial and astrophysical scenarios~\cite{Drewes:2013gca,Bolton:2019pcu,Abdullahi:2022jlv,Acero:2022wqg}, which places strong constraints on the parameter space of these heavy sterile neutrinos. 

\begin{figure}[t!]
    \includegraphics[width = \linewidth]{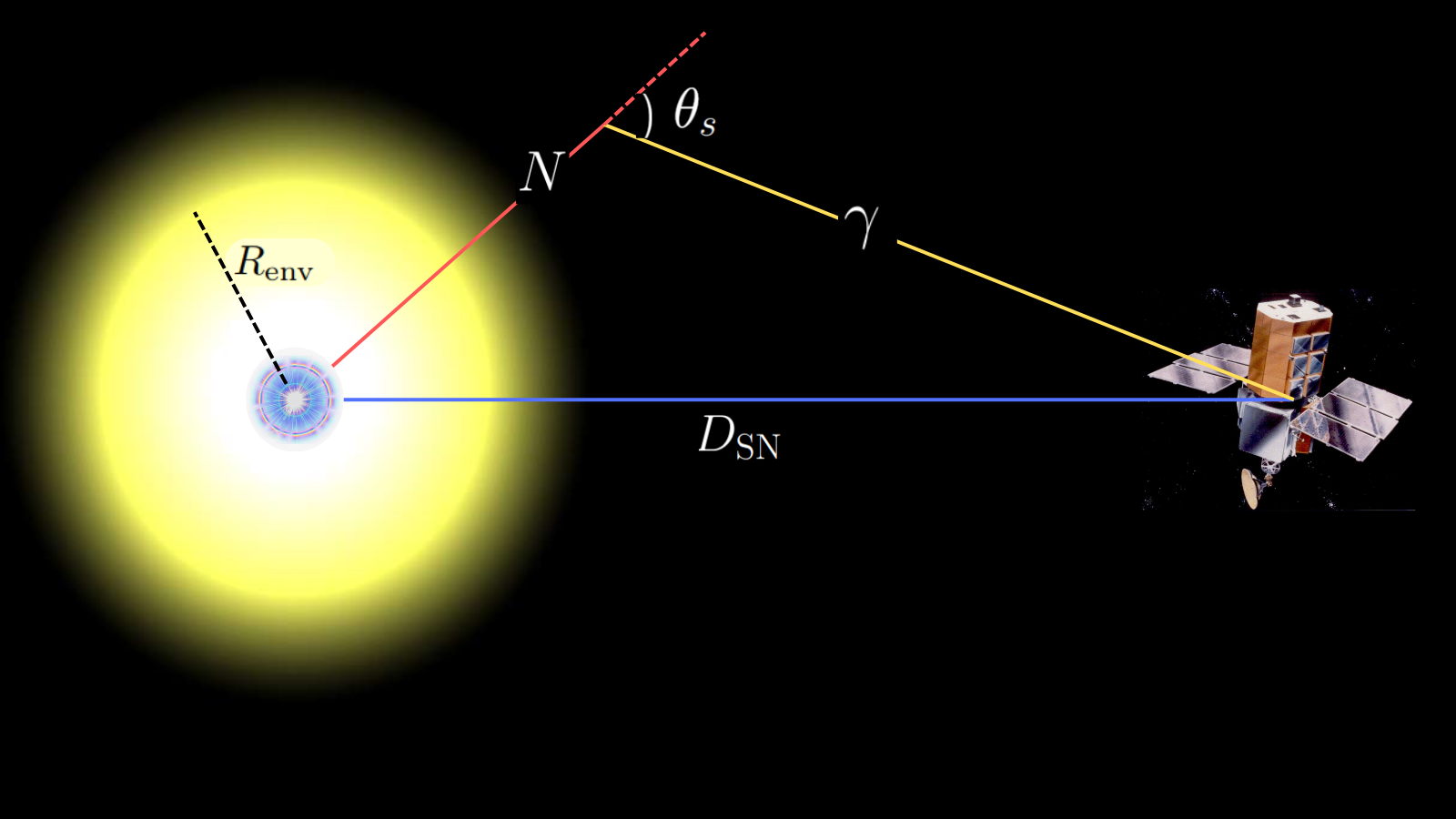}
    \caption{A pictorial representation of the sterile neutrino decay geometry (see text for details).}
    \label{fig:picgeom}
\end{figure}
A number of earlier works have studied the impact of heavy sterile neutrinos on supernovae through both mass-mixing~\cite{Falk:1978kf,Oberauer:1993yr,Dolgov:2000jw,Fuller:2008erj,Rembiasz:2018lok,Mastrototaro:2019vug,Carenza:2023old,Syvolap:2023trc,Chauhan:2023sci} and the dipole portal~\cite{Magill:2018jla,Brdar:2023tmi,Chauhan:2024nfa,Lazar:2024ovc}, while others have found strong constraints on light sterile neutrinos as well~\cite{Hidaka:2007se,Tamborra:2011is,Raffelt:2011nc,Suliga:2020vpz,Arguelles:2016uwb,Suliga:2019bsq,Syvolap:2019dat,Ray:2023gtu}. Although the SN1987A cooling bound is the most studied~\cite{Raffelt:1990yz}, the SN constraints which probe the smallest mixing angles arise from the non-observation of high-energy $\gamma$-rays at then existing spectrometers (e.g.~\cite{DeRocco:2019njg,Carenza:2023old}). With the upcoming and next-generation $\gamma$-ray telescopes such as Fermi-LAT and e-ASTROGAM, we aim to study the sensitivity of these experiments to the sterile neutrino parameter space. 

In this work, we also improve the $\gamma$-ray bound calculation for sterile neutrinos. Starting with the widely used general result for differential flux calculation in Ref.~\cite{Oberauer:1993yr} and simplifying, we find an extra condition on angular variable which theoretically affects results at longer times. We also explicitly include the $\Theta(x-R_{env})$ function to account for the SN envelope size to take into account that given an observation time $t_d$, there is a minimum boost factor (or $E_N$) that can reach the detector. Lastly, we find that for a diffuse $\gamma$-ray flux calculation, a factor of $1/4\pi$ should be accounted for the differential photon flux per steradian. In light of our improved calculation compared to the recent SN1987A $\gamma$-ray bounds, we also revisit the sterile neutrino bounds arising from Solar Maximum Mission (SMM) and the diffuse $\gamma$-ray background observations. 

This paper is organized as follows. In Sec.~\ref{sec:HNLprod}, we describe the production of sterile neutrinos inside the SN core. In Sec.~\ref{sec:NDecays}, we discuss the basic geometrical setup for the $\gamma$-ray flux calculation from the sterile neutrino decays. In Secs.~\ref{sec:gammarayobs} and~\ref{sec:diffusegammaray}, we show our results for bounds from SMM, diffuse $\gamma$-ray background and from observations of a nearby future galactic supernova. In Sec.~\ref{sec:discussion}, we discuss these results. Finally, we conclude in Sec.~\ref{sec:conclusion}.

\section{sterile neutrino Production}
\label{sec:HNLprod}
\noindent
The relevant Lagrangian for the sterile neutrino that couples to SM neutrinos through active-sterile mixing is

\begin{equation}
\mathcal{L} \, \supset \, i\bar{\nu}_4 \slashed{\partial}\nu_4 + y_{\nu} \bar{L} \tilde{H} \nu_4 - \frac{M_4}{2}\bar{\nu}_4^{c} \nu_4 + h.c.
\label{eq:lagr}
\end{equation}
with the first term being the kinetic term for $\nu_4$, second term is the Dirac mass term for neutrinos and the third term is the Majorana mass term for the sterile neutrino. Assuming mixing only with a single active neutrino flavor e.g. $\nu_\ell$ with $\ell=\mu,\tau$, then 
\begin{align}
     \nu_\ell & = U_{\ell 1}\, \nu_1 + U_{\ell 4} \, N \nonumber \\ 
     \nu_4 & = -U_{\ell 4} \, \nu_1 + U_{\ell 1} \, N      
\end{align}
where $U_{\ell i} = (y_\nu v M_4^{-1})_{\ell i}$ are the elements of the unitary mixing matrix with $v$ being the vacuum expectation value of the Higgs field. For small mixing angles $|U_{\ell 4}|\ll1$, $M_N\simeq M_4$, and since $U$ is unitary, $|U_{\ell 1}|^2=1-|U_{\ell 4}|^2$.  
where $|U_{\ell 4}|$ denotes the mixing between the sterile neutrino and the neutrino. 

Given the composition of the proto-neutron star, production primarily proceeds through scattering and annihilation processes involving (anti-)neutrinos ($\bar{\nu}_\ell/\nu_\ell$), electrons/positrons ($e^\pm$), muons ($\mu^-$), protons ($p$) and neutrons ($n$). All relevant production modes are listed in Table~\ref{tab:Nprod}. Until recently, the neutrino-neutrino scattering process was assumed to be the dominant mode. Recently, neutrino-nucleon scattering has been shown to be by far the most dominant production mode~\cite{Carenza:2023old}. We specifically find that neutrino-neutron scattering ($\nu_\ell+n\leftrightarrow n+N$) dominates over neutrino-proton scattering ($\nu_\ell+p\leftrightarrow p+N$). This difference can be easily attributed to the higher neutron number density compared to the protons.

\begin{table}[t]
    \centering
    \begin{tabular}{|c|}
    \hline
    Process \\
    \hline
$\nu_{\ell}+\bar{\nu}_{\ell}\leftrightarrow\bar{\nu}_{\ell}+N$  \\
$\nu_{\ell}+\nu_{\ell}\leftrightarrow\nu_{\ell}+N$ \\
$\nu_{\ell'}+\bar{\nu}_{\ell'}\leftrightarrow\bar{\nu}_{\ell}+N$ \\
$\nu_{\ell}+\bar{\nu}_{\ell'}\leftrightarrow\bar{\nu}_{\ell'}+N$  \\
$\nu_{\ell}+\nu_{\ell'}\leftrightarrow\nu_{\ell'}+N$ \\
$e^++e^-\leftrightarrow\bar{\nu}_{\ell}+N$ \\
$\nu_{\ell}+e^-\leftrightarrow e^-+N$ \\
$\nu_{\ell}+e^+\leftrightarrow e^++N$ \\
$\nu_\ell+n\leftrightarrow n+N$ \\
$\nu_\ell+p\leftrightarrow p+N$ \\
$\mu^-+p\leftrightarrow n+N$ \\
$\mu^-+\nu_e\leftrightarrow e^-+N$ \\
    \hline
    \end{tabular}    
\caption{The production modes for sterile neutrinos inside the proto-neutron star core~\cite{Carenza:2023old} (for matrix elements, see App.~\ref{app:decaywidths}).} 
\label{tab:Nprod}
\end{table}
The sterile neutrino production calculation in a hot and dense medium requires solving the Boltzmann transport equations. Under the assumptions of isotropic and homogeneous environment, the expression for the differential number flux can be written as 
\begin{equation}
    \frac{d^2N_N}{dE_N dt} = \frac{1}{2\pi^2} \int d^3r\, \frac{df_N}{dt} E_N\,p_N.
\label{eq:diff} 
\end{equation}
where the rate change in sterile neutrino phase space can be expressed by the following $2\leftrightarrow2$ collisional integral
\begin{align}
    \frac{\partial f_N}{\partial t} &= \frac{1}{2 E_N} \int \frac{d^3 p_1}{(2 \pi^3)\,2 E_1} \frac{d^3 p_2}{(2 \pi^3)\,2 E_2} \frac{d^3 p_3}{(2 \pi^3)\,2 E_3}\, |M|^2_{N3\rightarrow 12}  \nonumber \\ & \times \Lambda(f_1,f_2,f_3,f_N) \, \delta^4(p_N+p_3-p_1-p_2) (2 \pi)^4,
    \label{eq:collint}
\end{align}
where $\Lambda(f_1,f_2,f_3,f_N)= (1-f_N)(1-f_3)f_1 f_2- f_N f_3(1-f_1)(1-f_2)$ is the phase-space factor including the Pauli blocking of final states, $|M|^2$ is the interaction matrix element element squared including the symmetry factor. For small mixing angles of interest, we expect the sterile neutrino not to be thermalized, therefore we set $f_N=0$.

Under the high temperature and pressure in a SN core, the dense nuclear matter leads to a breakdown of the non-interacting picture. The nucleon self-energies arising from strong interactions are modeled as a non-relativistic quasi-particle gas moving in a mean-field potential~\cite{Martinez-Pinedo:2012eaj,Mirizzi:2015eza}. This leads to a modified dispersion relation for which the nucleon phase-space distribution can be expressed in a similar form to the standard Fermi-Dirac distribution (as for leptons) 
\begin{equation}
    f_{\text{nucleon}}(p)=\frac{1}{\text{exp}\left[\frac{\sqrt{p^2+{m^*}^2}-\mu^*}{T} \right]+1},
\end{equation}
where we define the effective nucleon chemical potential $\mu^*=\mu-U$ with $\mu$ being the nucleon chemical potential (with rest mass included) and $m^*$ is the Landau effective mass of the nucleon.
In this work, we use the 18.8 $M_\odot$ simulated SN by the Garching group (SFHo-18.8) assuming $R_{\text{core}}\sim 20$ km and a stellar envelope $R_{\text{env}} \sim 5 \times 10^{8}$ km for all post-bounce time sequences up to $10$ s \cite{SNprofile,Mirizzi:2015eza,Bollig:2020xdr}. 

We note that the sterile neutrino production rate to be used in Sec.~\ref{sec:Geom} (see Eq.~\ref{eq:dNdEdt}) depends on two primary assumptions, (a) The emission arises from the center of the SN core, i.e. point emission, (b) The emission is instantaneous i.e. $t_d=0$. The first assumption is justified since we are concerned with the parameter space when the sterile neutrino mean free path length is larger than the SN envelope size. Therefore, the sub-leading correction factors are expected to be $\mathcal{O}(R_{\text{core}}/R_{\text{env}})$, where $(R_{\text{core}}/R_{\text{env}})\ll 1$. For the second assumption, note that the major portion of the sterile neutrino production occurs primarily before the core cools down substantially which is usually under 3-4 seconds. Since the observation timescales for $\gamma$-ray telescopes are usually far longer, the constraints arising from these measurements are not affected. Therefore under these two assumptions, we can express the total differential number spectrum by integrating over the entire volume upto $r=R_{\text{core}}$ and for times up to the entire duration of the simulated SN profile $\sim$ 10 s.

Finally, we also include the effects of gravitational red-shifting by including the lapse factor $\eta_{\text{lapse}}(r)$ and for gravitational trapping by including a sharp cut-off in sterile neutrino spectrum below $E_N = \eta_{\text{lapse}}^{-1} M_N$~\cite{Chauhan:2024nfa}. In addition, The resulting differential neutrino spectrum is given by 
\begin{equation}
    \frac{dN_N}{dE_N} = \int dt \int \frac{d^3r}{2\pi^2} \,\eta_{\text{lapse}}^2(r) \frac{df_N}{dt} E_N\,p_N\, \Theta\left(E_N -\frac{M_N}{\eta_{\text{lapse}}} \right) 
\label{eq:diffFinal} 
\end{equation}

For integration up to SN core, we need to take into effect the absorption effects. The averaged absorption factor including the non-radial propagation effects which is given by ~\cite{Caputo:2021rux}
\begin{equation}
    \langle e^{-\tau(E_N,r)} \rangle = \frac{1}{2}\int_{-1}^{+1} d\omega~ e^{-\int_0^{\infty} ds\, \Gamma_{abs}\left(E_N,\sqrt{r^2+s^2+2rs\omega}\right)}
    \label{eq:absorb}
\end{equation}
where $\omega \equiv \cos \theta$, and the absorption rate $\Gamma_{abs}$ has contributions from both scatterings and decays. The absorption rate $\Gamma_{abs}$ is given by an expression similar to the collisional term~\cite{Weldon:1983jn},
\begin{align}
     \Gamma_{abs} = &\frac{1}{2 p_N} \int \frac{d^3 p_1}{(2 \pi^3)\,2 E_1} \frac{d^3 p_2}{(2 \pi^3)\,2 E_2} \frac{d^3 p_3}{(2 \pi^3)\,2 E_3}\, \Tilde{\Lambda}(f_1,f_2,f_3) \nonumber \\ &   \times |M|^2_{N3\rightarrow 12}\, \delta^4(p_N+p_3-p_1-p_2) (2 \pi)^4,
\end{align}
where $\Tilde{\Lambda}(f_1,f_2,f_3)= f_3(1-f_1)(1-f_2)$. Qualitatively in the trapping regime, the couplings are larger and therefore production regions with higher absorption rates get suppressed in the energy integral Eq.~\eqref{eq:absorb}. Therefore, the dominant contribution arises from regions with the least absorption rates, i.e., regions near $R_{\rm core}$. In these outer regions near the core, the proton and electron number density is comparatively lower, therefore the absorption rate is dominated by the decays of $N$, which sets the maximum allowed coupling strength in the trapping regime.

Note that at this stage we do not include an exponential factor for attenuation effects from absorption effects outside the SN core but inside the SN envelope, since this would be taken into account explicitly for the $\gamma$-ray calculation in Eq.~\eqref{eq:dNdEdt}. The important fact for $\gamma$-ray bounds which is primarily concerned with propagation lengths $\mathcal{O}(R_{\text{env}})$ (far greater than $R_{\text{core}}$) is that only decay modes rather than scattering processes play a dominant role for absorption 
effects~\cite{Chauhan:2024nfa}. 

\section{sterile neutrino Decays \label{sec:NDecays}}
\subsection{Decay Geometry \label{sec:Geom}}
The geometrical picture for $\gamma$-ray production through sterile neutrino decays proceeds as follows : In the instantaneous and point source emission of sterile neutrino (assumed to be isotropic) at $r=0$, the sterile neutrino propagates for a distance $x$ before decaying and emits a photon at angle $\theta_{s}$ relative to the sterile neutrino propagation direction as measured in the supernova/lab frame (see Fig.~\ref{fig:picgeom}). For the time being, we will consider all neutrinos and sterile neutrinos to be produced at the same instant, and define the time delay $t_{d}$ to be the difference in the times of arrival between the first neutrino event and any subsequent photons. The time delay is given by
\begin{equation}
    t_{d} = x \bigg( \frac{1}{\beta} - \cos \theta_{s} \bigg)
    \label{eq:Time_delay_from_x}
\end{equation}
where $\beta = (1 - M_N^2/E_N^2)^{1/2}$. Following the work in Ref.~\cite{Oberauer:1993yr}, we can relate the SN/lab frame decay angle and photon energy ($\theta_{s}$, $E_{\gamma}$) to the values in the sterile neutrino rest frame ($\theta_{N}$,$\omega$) by
\begin{equation}
    \cos \theta_{s} = \frac{\beta + \cos \theta_{N}}{1 + \beta \cos \theta_{N}},
    \label{eq:Cos_Lab}
\end{equation}
\begin{equation}
    E_{\gamma} = \gamma \omega ( 1 + \beta \cos \theta_{N}),
    \label{eq:Energy_Lab}
\end{equation}
where $\gamma = E_{N}/M_{N}$. For now, we will assume the decays to be isotropic in the sterile neutrino rest frame and let $g_{N}(\omega)$ denote the energy distribution for the photons in this frame. 

We can relate the rest-frame distribution to the energy distribution in the supernova frame $g_{s}(E_{\gamma})$ via
\begin{equation}
    g_{s}(E_{\gamma}) = g_{N} \bigg(\frac{E_{\gamma}}{\gamma (1+ \beta \cos \theta_{N})} \bigg) \frac{1}{\gamma (1+ \beta \cos\theta_{N})} 
\end{equation}
Turning now to sterile neutrino decay, we know that
\begin{equation}
    \frac{1}{\lambda} \exp(-x/\lambda) dx = P(t_{d}) dt_{d}
\end{equation}
where $P(t_{d})$ is the probability density for decay in terms of the measured time delay. Using Eqs. \eqref{eq:Time_delay_from_x} and \eqref{eq:Cos_Lab}, we can write
\begin{equation}
    x = t_{d} \gamma^2 \beta (1 + \beta \cos \theta_{N}).
    \label{eq:distx}
\end{equation}
As we know the decay length $\lambda = \gamma \beta \tau$, we can now determine that
\begin{equation}
    P(t_{d}) = \frac{\gamma (1+ \beta \cos\theta_{N})}{\tau} \exp\bigg(\frac{-\gamma t_{d} (1+\beta \cos \theta_{N})}{\tau} \bigg).
\end{equation}
Putting everything together, we can find the differential rate of photons to be
\begin{equation}
    \begin{split}
    \frac{d^2 N_{\gamma}}{dE_{\gamma} dt_{d}} &= \int dE_{N} d\cos\theta_{N} \frac{g_{s}(E_{\gamma}) P(t_{d})}{2} \frac{dN_{N}}{dE_{N}} \Theta(x - R_{env}) \\
    &= \int dE_{N} d\cos\theta_{N} \frac{1}{2\tau} g_{N} \bigg(\frac{E_{\gamma}}{\gamma (1 + \beta \cos \theta_{N})} \bigg)\\
    &\times \exp \bigg(\frac{- \gamma t_{d} (1 + \beta \cos \theta_{N})}{\tau} \bigg) \frac{dN_{N}}{dE_{N}} \Theta(x - R_{env}),
    \end{split}
    \label{eq:dNdEdt}
\end{equation}
where $\Theta(y)$ is the Heaviside theta function and $\frac{dN_{N}}{dE_{N}}$ is given by expression in Eq.~\eqref{eq:diffFinal}. To convert the above expression to flux, we divide by $4 \pi D_{SN}^2$, and we integrate over the desired time and energy window to obtain a fluence. See App.~\ref{app:GenExp} for possible simplifications.

\begin{figure}[t]
    \includegraphics[width = \linewidth]{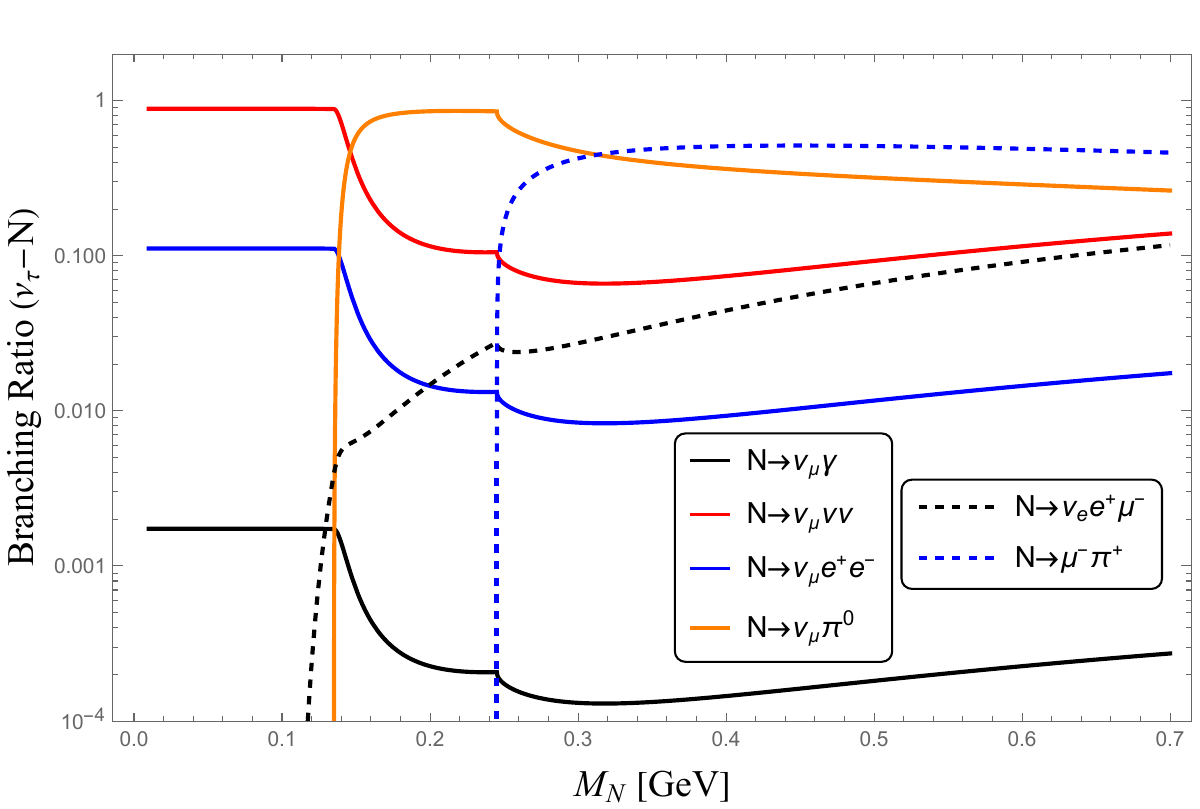}
    \includegraphics[width = \linewidth]{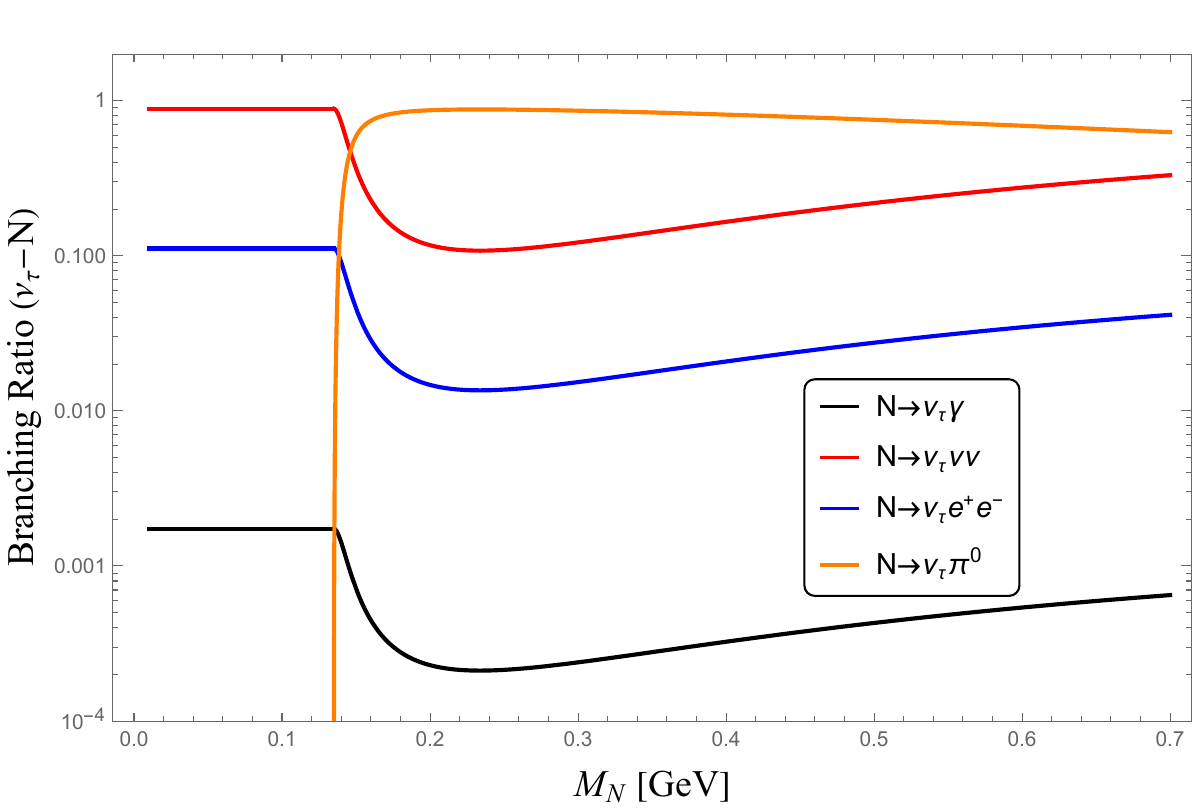}
    \caption{Branching ratios for the decays of $N$ for coupling exclusively to muon (\textbf{top}) or tau (\textbf{bottom}) neutrinos. We use the decay widths given in Ref.~\cite{Coloma:2020lgy} and ignore the subdominant $N \rightarrow \nu \mu^{-} \mu^{+}$ decays. \label{fig:branching-ratios}}
\end{figure}

\begin{figure*}[ht!]
    \subfloat{
      \includegraphics[width=0.45\textwidth]{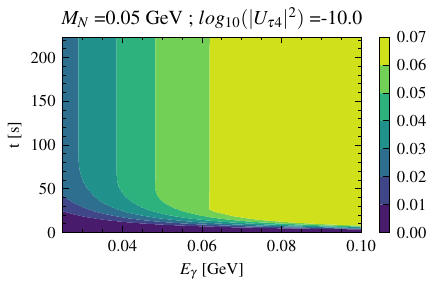}
    }
    \hfill
    \subfloat{
      \includegraphics[width=0.45\textwidth]{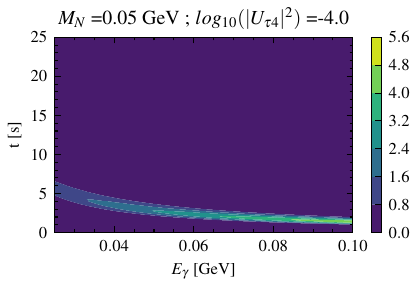}
    }
    \\
        \subfloat{
      \includegraphics[width=0.45\textwidth]{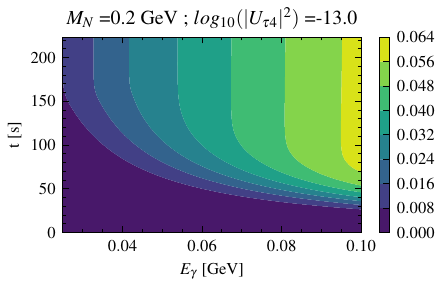}
    }
    \hfill
    \subfloat{
      \includegraphics[width=0.45\textwidth]{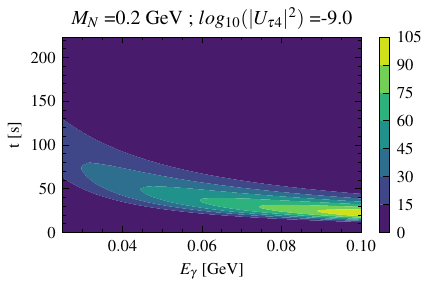}
    }
    \caption{Differential flux $\frac{d^2\Phi_{\gamma}}{d E_{\gamma} dt}$ in $\mathrm{cm^{-2} s^{-1} GeV^{-1}}$ from SN1987A for characteristic parameter points A-D (see text) and $R_{env} = 2 \times 10^{12}$ cm. The time and energy windows are chosen to align with the SMM (with the exception of Point B, which is zoomed in to see the details). \label{fig-Char-Points}}
\end{figure*}

\subsection{Decay Rates}
The sterile neutrino decay rates have been calculated in the literature~\cite{Gorbunov:2007ak, Atre:2009rg, Bondarenko:2018ptm,Coloma:2020lgy,Helo:2010cw}. In this work, we have used the decay rate expressions from Ref.~\cite{Coloma:2020lgy}, provided in App.~\ref{app:decaywidths}. Note that the expression for $\pi^0$ decay mode in Refs.~\cite{Gorbunov:2007ak,  Bondarenko:2018ptm,Coloma:2020lgy} differ by a factor of $2$ (in the numerator) compared to Refs~\cite{ Atre:2009rg,Helo:2010cw}. The branching ratios for the sterile neutrino coupling with $\mu$ and $\tau$ flavors are shown in Fig.~\ref{fig:branching-ratios}. Note that the branching ratios do not depend on the mixing angles. 

The differential photon flux as derived in Eq.~\eqref{eq:dNdEdt} depends on the normalized decay distribution $g_N(\omega)$ as a function of photon energy $\omega$ in the sterile neutrino rest frame. We discuss the distribution function for the two relevant decay modes below:

\subsubsection{ N $\rightarrow \gamma \nu$}
For the two-body decay of $N \rightarrow \gamma \nu$, we know the decay is isotropic and energy spectrum is a Dirac delta distribution in the sterile neutrino frame. Therefore, the normalized decay distribution for this decay mode is 
\begin{equation}
    g_{N \rightarrow \gamma \nu}(\omega) = B_{\gamma \nu} \delta(\omega - M_N/2),
\end{equation}
where $B_{\gamma \nu}$ is the branching fraction for the decay mode. Below the pion mass threshold, this decay is the only mode leading to photon production. Using the above $g_{N \rightarrow \gamma \nu}(\omega)$ in Eq.~\eqref{eq:dNdEdt} and integrating over $\cos\theta_N$, we obtain a differential rate of
\begin{equation}
    \begin{split}
    \frac{d^2 N_{\gamma}}{dE_{\gamma} dt_{d}} =& \int dE_{N} \frac{2 E_{\gamma} B_{\gamma \nu}}{\gamma m^2_{N} \beta \tau}
    \exp \bigg(\frac{- 2 E_{\gamma} t_{d}}{\tau M_N} \bigg) \frac{dN_{N}}{dE_{N}} \\
    &\Theta(x - R_{env}) \Theta \bigg( E_{\gamma} - \frac{E_{N} (1 - \beta)}{2} \bigg) \\
    &\Theta \bigg( \frac{E_{N} (1 + \beta)}{2} - E_{\gamma} \bigg).
    \end{split}
    \label{eq:NnuGMain}
\end{equation}
where last two $\Theta(...)$ functions arise from the requirement $|\cos\theta_N|\leq 1$. This can be further simplified (see App.~\ref{app:Ngammanu} for details) and can be written as 
\begin{equation}
    \begin{split}
    \frac{d^2 N_{\gamma}}{dE_{\gamma} dt_{d}} = & \frac{2 E_{\gamma} B_{\gamma \nu}}{ M_N \tau} 
    \exp \bigg(\frac{- 2 E_{\gamma} t_{d}}{\tau M_N} \bigg) \\ & \times \int_{E'_{N,\text{min}}}^\infty dE_{N} \frac{1}{p_N} \frac{dN_{N}}{dE_{N}}
    \end{split}
\end{equation}
where $E'_{N,\text{min}}=\text{Max}[E_{N,\text{min}}^{1,2},M_N]$ with
\begin{equation}
    E_{N,\text{min}}^1 =\frac{E_\gamma^2+\left(\frac{M_N}{2}\right)^2}{E_\gamma},\,  E_{N,\text{min}}^2 = M_N\sqrt{1+\left(\frac{ R_{env} M_N}{2 t_d E_\gamma}\right)^2}
    \label{eq:Emins}
\end{equation}
We note that the dependence on several variables for the differential flux in a previous study~\cite{Carenza:2023old} was completely flipped. Our results also closely agree with original derivation in Ref.~\cite{Oberauer:1993yr} except the added condition for $E_{N,\text{min}}^1$, which arises from ensuring $|\cos\theta_N|\leq 1$. This condition was missed in the original derivation (see App.~\ref{app:Ngammanu}). 

\subsubsection{N $\rightarrow \pi^{0} \nu$}
Above the $\pi^0$ mass threshold,  $N \rightarrow \pi^{0} \nu$ decay mode becomes kinematically accessible. Compared to the $N \rightarrow \gamma \nu$, the branching ratio for $\pi^0$ mode is nearly $1$, while for the former is $\mathcal{O}(10^{-3})$. In the decay of $N$ through $\pi^{0} \nu$ is followed almost immediately by $\pi^{0} \rightarrow \gamma \gamma$. In the sterile neutrino rest frame, the pion will be produced mono-energetically at
\begin{equation}
    E_{\pi} = \frac{M_N^2 + m_{\pi}^2}{2 M_N}.
\end{equation}
This will allow us to define $\gamma_{\pi} = E_{\pi}/m_{\pi}$ and $\beta_{\pi} = \sqrt{1 - 1/\gamma_{\pi}^2}$. The pion decay will be a 2-body decay into massless particles similar to the previous case for $N\rightarrow \gamma \nu$. Therefore, the normalized energy distribution takes the form of box distribution 
\begin{equation}
    \begin{split}
    g_{N \rightarrow \pi \nu}(\omega) = \frac{2 B_{\pi \nu}}{\gamma_{\pi} \beta_{\pi} m_{\pi}} \Theta \bigg( \omega - \frac{E_{\pi}(1 - \beta_{\pi})}{2} \bigg) \\ 
    \Theta \bigg( \frac{E_{\pi}(1 + \beta_{\pi})}{2} - \omega \bigg),
    \end{split}
\end{equation}
which we can then include in Eq.~\eqref{eq:dNdEdt} and integrate numerically.

\begin{figure*}[ht!]
    \subfloat{
      \includegraphics[width=0.45\textwidth]{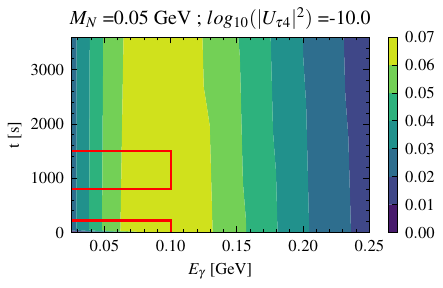}
    }
    \hfill
    \subfloat{
      \includegraphics[width=0.45\textwidth]{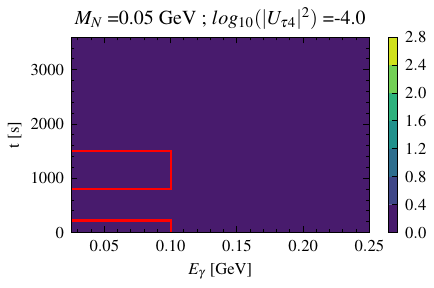}
    }
    \\
        \subfloat{
      \includegraphics[width=0.45\textwidth]{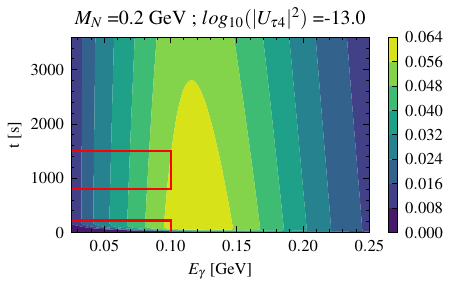}
    }
    \hfill
    \subfloat{
      \includegraphics[width=0.45\textwidth]{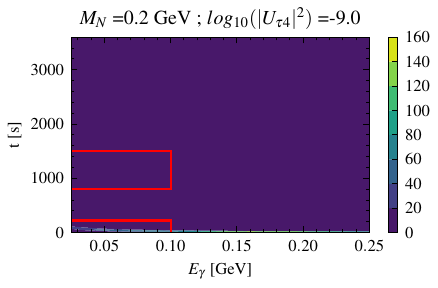}
    }
    \caption{Differential flux $\frac{d^2\Phi_{\gamma}}{d E_{\gamma} dt}$ in $\mathrm{cm^{-2} s^{-1} GeV^{-1}}$ from SN1987A for characteristic parameter points A-D (see text) and $R_{env} = 2 \times 10^{12}$ cm. We have extended the time and energy windows, with red boxes to indicate the observations of the SMM. The first box extends from 0 to 223.2 seconds. The second time window is from 800 to 1500 seconds, which indicates the late-time observations of the SMM after its calibration time. \label{fig-Char-Points-Extended}}
\end{figure*}

\section{Gamma Ray Observations}
\label{sec:gammarayobs}
To demonstrate the qualitative behavior of the $\gamma$-ray flux, we will consider four example points in parameter space.
\begin{itemize}
    \item A: $M_N$ = 50 MeV, $|U_{\ell 4}|^2 = 10^{-10}$
    \item B: $M_N$ = 50 MeV, $|U_{\ell 4}|^2 = 10^{-4}$
    \item C: $M_N$ = 200 MeV, $|U_{\ell 4}|^2 = 10^{-13}$
    \item D: $M_N$ = 200 MeV, $|U_{\ell 4}|^2 = 10^{-9}$
\end{itemize}
These mass values are chosen so that they lie above and below the $N \rightarrow \pi^{0} \nu$ threshold, and the $|U|^{2}$ values are near the upper and lower exclusions possible with SN1987A. We can see the predicted differential flux for these parameter points during the 223s after the first neutrino arrival in Fig. \ref{fig-Char-Points}. We note that all parameters points have the largest flux near the cutoff photon energy of 100 MeV. Next, we see that high energy photons appear before lower energies, as lower energies require either larger decay angles, or less energetic parent $N$, both of which increase the time of delay. We see that for smaller couplings (longer decay lengths), there is a period where the flux is roughly time-independent. For larger couplings (shorter decay lengths), the flux quickly falls away as the $N$ quickly decay.

Observing these characteristic parameter points indicate how future observations may improve discovery potential (see Fig. \ref{fig-Char-Points-Extended}). First, extending the energy range allows for more of the flux to be observable. For smaller couplings especially, we can see that the flux remains sizable for several thousand seconds, so increasing exposure time would extend the parameter space able to be probed. Moreover, as the photon flux has a different energy and time distribution depending on $M_N$ and $|U|^2$, if an excess of $\gamma$-rays is seen in the next galactic supernova, the spectrum could imply the properties of the sterile neutrino. Estimating the precision at which these properties could be determined is a non-trivial task, and we delay that calculation to future work.

\begin{figure}
    \includegraphics[width = \linewidth]{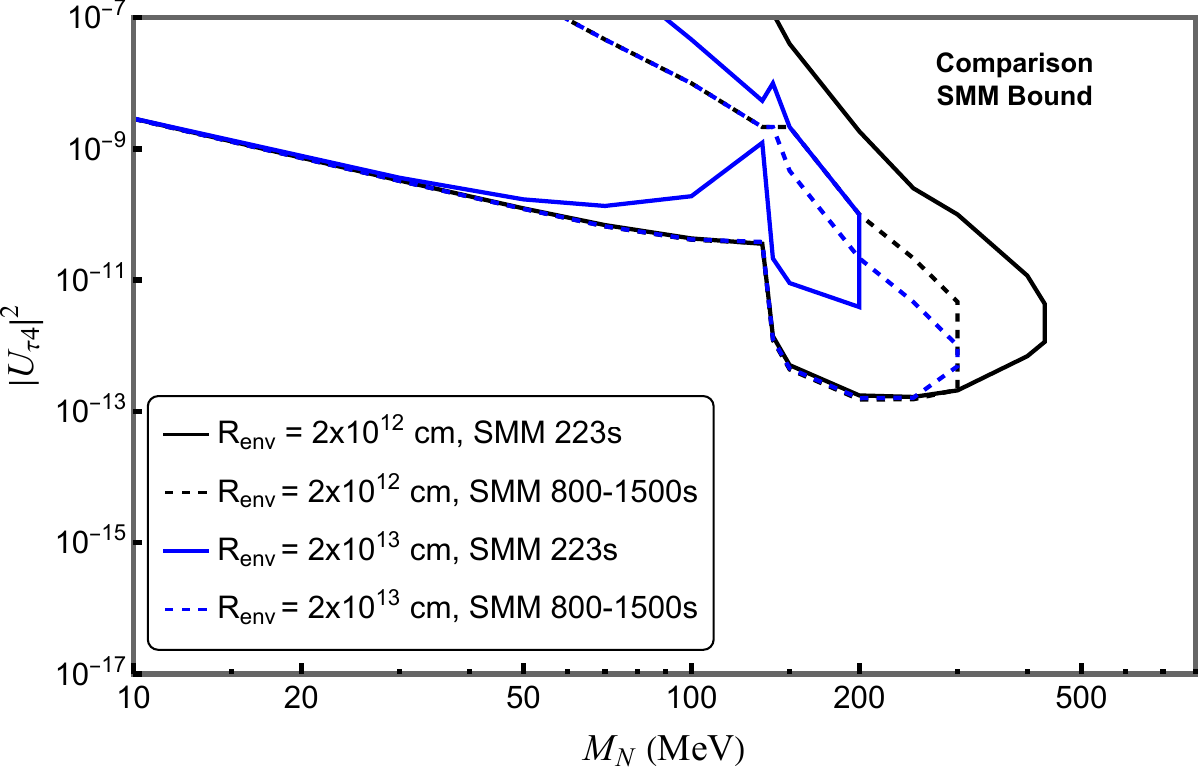}
    \caption{Constraints on $|U_{\tau, 4}|^2$ from observations of the SMM at early (0-223s) and late (800-1500s) times. We repeat this for different guesses of the supernova envelope size, seeing that a larger envelope means that later times are able to place the stronger constraint.  \label{fig-Early-and-Late}}
\end{figure}

\begin{figure*}
    \includegraphics[width = \linewidth]{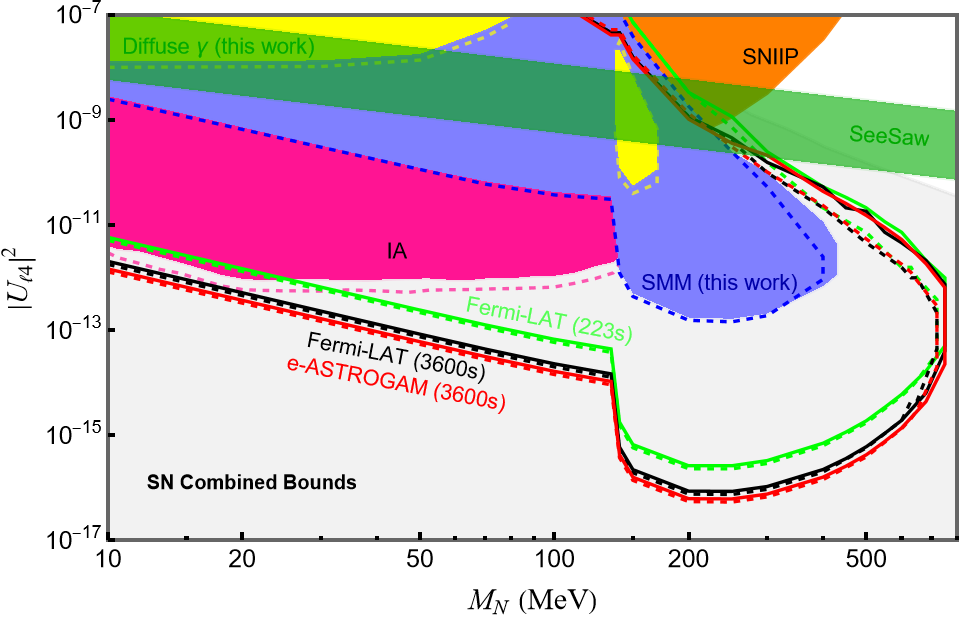}
    \caption{The combined current bounds on $|U_{\ell 4}|^2$ derived in our work (solid curves for $|U_{\tau 4}|^2$ and dotted for $|U_{\mu 4}|^2$ ) from SN1987A with SMM observations and diffuse $\gamma$-ray background along with future exclusion projections assuming a future galactic supernova 10 kpc from Earth with Fermi-LAT and e-ASTROGAM observations. Other SN bounds shown are low-energy SNIIP~\cite{Chauhan:2023sci,Carenza:2023old} and in-flight positron annihilation (IA) bound~\cite{DelaTorreLuque:2024zsr}. The light gray shaded region is bound from BBN assuming standard cosmology~\cite{Boyarsky:2009ix,Ruchayskiy:2012si,Sabti:2020yrt,Domcke:2020ety} and the parameter region favored by conventional Type-I seesaw models labeled as `SeeSaw'. \label{fig:U2-Constraints}}
\end{figure*}

\begin{figure*}[ht!]
    \subfloat{
      \includegraphics[width=0.45\textwidth]{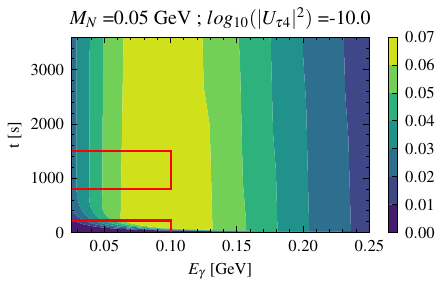}
    }
    \hfill
    \subfloat{
      \includegraphics[width=0.45\textwidth]{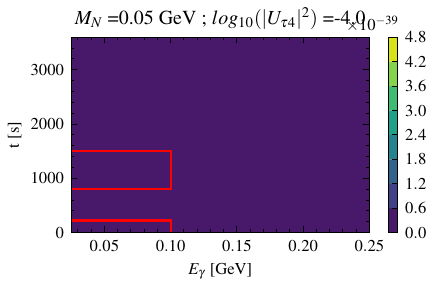}
    }
    \\
        \subfloat{
      \includegraphics[width=0.45\textwidth]{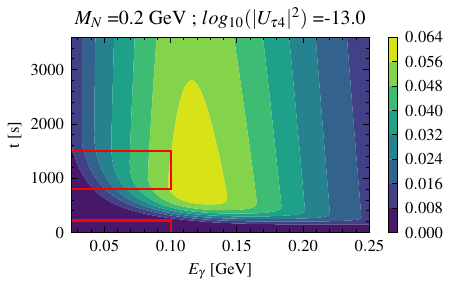}
    }
    \hfill
    \subfloat{
      \includegraphics[width=0.45\textwidth]{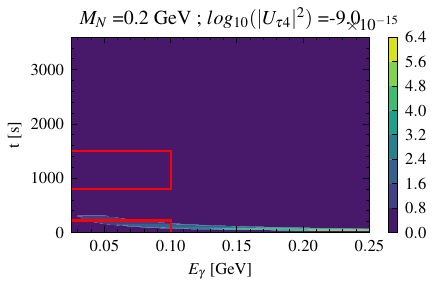}
    }
    \caption{Same differential flux graphs as Fig.~\ref{fig-Char-Points-Extended}, except with a $R_{env} = 2 \times 10^{13}$ cm. We note that for lower couplings, the differential flux is pushed towards later times and higher energies. For the larger couplings, we see that the flux has decreased by several orders of magnitude. However, within the late time observations (800s-1500s), the results are comparable to Fig.~\ref{fig-Char-Points-Extended}. We can see the effects of such a progenitor and choice of observation time window in Fig.~\ref{fig-Early-and-Late}, where the high-coupling and high-mass constraints become weaker as the envelope size increases for early time observations. However, late time observations are less sensitive to envelope size. \label{fig-Large-Envelope-Char-Points}}
\end{figure*}
Taking these calculations for the $\gamma$-ray flux, we can place limits on the matrix element coupling the sterile neutrino to active neutrinos. For photons produced from decaying sterile neutrinos during SN1987A, they could have been detected by the Gamma-Ray Spectrometer (GRS) on the Solar Maximum Mission (SMM). However, there was no excess of $\gamma$-ray events detected in conjunction with the neutrino burst. Using the non-observation, the 2-$\sigma$ upper bound on the fluence of $\gamma$-rays in the energy range 25-100 MeV for a duration of 223.2s was set to be $1.38\,\mathrm{cm}^{-2}$~\cite{Chupp:1989kx}. After the initial interval of 223.2s after the first detected neutrino event, the SMM went into the calibration mode for about 10 minutes. In this work, we will also for the first time explore the constraints arising from the second observation window after the calibration period, which we estimate to be 800-1500 seconds after the first neutrino arrival. Most previous works \cite{Brdar:2023tmi,Carenza:2023old,Oberauer:1993yr,Chupp:1989kx} use the time period prior to calibration to set $\gamma$-ray constraints.

Unlike pre-calibration times where we have detailed data on the $\gamma$-ray observations \cite{Chupp:1989kx}, we do not have numeric values on the post-calibration observations. To perform our analysis, we use the same flux limit as the first 223 seconds; that is to say that over an observation time of $\Delta t$, we exclude BSM fluences above $1.38 \times (\frac{\Delta t}{223 \mathrm{s}}) \mathrm{cm^2}$.

As we can see in Fig.~\eqref{fig-Early-and-Late}, observations at later times are less effective at constraining large values of $|U_{\tau 4}|^2$. At such large couplings, most of the decays occur at early times, so the flux has drastically diminished by post-calibration times. For a supernova envelope size of $2 \times 10^{12}$ cm, the bounds on small couplings and masses below 300 MeV are identical between early- and late-time measurements. This is because 223s is sufficient time for many of the sterile neutrinos to exit an envelope of this size, and at such couplings they are long-lived so the photon flux extends past 1500s. However, for a larger envelope of $2 \times 10^{13}$ cm, we see that late-time observations are better, especially for larger masses as most of the sterile neutrino flux is unable to exit the large envelope in 223s as shown for our characteristic parameter points in Fig \ref{fig-Large-Envelope-Char-Points}. This indicates the importance of late-time measurements for supernova with large, optically-thick envelopes.

We also study the sensitivity of current and near-future $\gamma$-ray observatories capable of probing a larger sterile neutrino parameter space, especially for very low mixing angles. For this purpose, we will consider a future galactic SN at a distance $D_{\text{SN}}=10$ kpc. We will consider the existing Fermi-LAT space observatory with an effective area of 9500 $\mathrm{cm^{2}}$, an angular resolution of $5^{\circ}$, which can detect photon with energies between 20 MeV - 300 GeV~\cite{FermiLAT}. For a near-future mission, we will consider the proposed e-ASTROGAM observatory by the European Space Agency, with a focus to cover the MeV $\gamma$-ray band~\cite{e-ASTROGAM:2017pxr}. e-ASTROGAM will have an effective detection area of 9025 $\mathrm{cm^2}$, angular resolution of $1.25^{\circ}$ and detection sensitivity in 0.3 MeV - 3 GeV. 

For our analysis, we will restrict $E_\gamma>100$ MeV for both Fermi-LAT and for e-ASTROGAM. The extragalactic $\gamma$-ray background (EGB) forms the primary background in our case and has been well measured by COMPTEL~\cite{COMPTEL:1994a,COMPTEL:2000a}, EGRET~\cite{EGRET:1997qcq,Strong:2004ry} and Fermi-LAT~\cite{Fermi-LAT:2014ryh}\footnote{We note that the Fermi-LAT measurements are lower than EGRET, due to the superior sensitivity of the former to resolve individual sources, leading to a better foreground subtraction.}. We define the log-likelihood as~\cite{Brdar:2023tmi}
\begin{equation}
    -2\ln{\mathcal{L}} = 2 \left(N_{\text{exp}}-N_{\text{obs}} + N_{\text{obs}}\text{Log}\left[\frac{N_{\text{obs}}}{N_{\text{exp}}}\right]\right)
\end{equation}
and study the future constraints at the $2\sigma$-level (i.e. $-2\ln{\mathcal{L}}= 3.841$). Here $N_{\text{obs}}$ denotes number of observed events, which is assumed to be consistent with the background expectation to set constraints. The expected EGB background flux for $E_\gamma>100$ MeV is around $1.5 \times 10^{-5} \text{ cm}^{-2}\text{s}^{-1}\text{sr }$~\cite{Jaeckel:2017tud}. Using this background photon flux, limits for Fermi-LAT can be set at 2.4 events over 223s and 4.8 events over 3600s.  For e-ASTROGAM, limits can be set at 2.4 events over 3600s  for $E_\gamma>100$ MeV. We show the projection of these constraints in Fig.~\ref{fig:U2-Constraints} (see discussion in Sec.~\ref{sec:discussion}). We do not study the the e-ASTROGAM case for $E_\gamma>1$ MeV, since the expected EGB background flux is quite high around $4.2 \times 10^{-3} \text{ cm}^{-2}\text{s}^{-1}\text{sr}$. In such a case, limits can only be set at 15.4 events over 3600s and the resulting constraints are weaker than when we limit to $E_{\gamma} > 100$ MeV.

\section{Diffuse $\gamma$-Ray Flux}
\label{sec:diffusegammaray}
The radiative sterile neutrino decays arising from all past SNe in the history of the universe will contribute to the diffuse $\gamma$-ray background. We can also place constraints by requiring this background to be below the observed diffused flux of extragalactic $\gamma$-rays from Fermi-LAT. The latest measurements from Fermi-LAT for the diffuse background can be described by the following power law~\cite{Fermi-LAT:2014ryh,Calore:2020tjw} 
\begin{equation}
    \frac{d \Phi_{\gamma}^{\text{\tiny{Fermi}}}}{dE_{\gamma}} = 2.2 \times 10^{-3} (E_{\gamma}/\mathrm{MeV})^{-2.2} \mathrm{cm^{-2}} \mathrm{s^{-1}} \mathrm{MeV^{-1}} \mathrm{sr^{-1}}
    \label{eq:fermiLat}
\end{equation} 

In analogy to the diffuse supernova neutrino background, the diffuse $\gamma$-ray flux from sterile neutrino decays can be written as~\cite{Lunardini:2009ya, Beacom:2010kk}
\begin{equation}
    \frac{d \Phi_{\gamma}^N}{dE_{\gamma}} = \frac{1}{4\pi}\int dz (1+z) \frac{d N_{\gamma}(E_{\gamma}(1+z))}{dE_{\gamma}} R_{SN}(z) \frac{d \ell}{dz}
    \label{eq-Diffuse-Flux}
\end{equation}
where the differential distance $({d \ell}/{dz})$ is
\begin{equation}
    \frac{d \ell}{dz} = \frac{c}{H_{0} (1+z) [\Omega_{\Lambda} + \Omega_{M}(1+z)^{3}]^{1/2}}
\end{equation}
and the core-collapse supernova rate density $R_{SN}(z)\simeq R_{SF}(z)/(71\,M_\odot)$ follows the star formation rate $R_{SF}$ \cite{SNRate}
\begin{equation}
    R_{SF} \propto 
    \begin{cases}
        (1+z)^{\beta} \, \, 0<z<1 \\
        2^{\beta - \alpha} (1+z)^{\alpha} \, \, 1<z<4.5 \\
        2^{\beta-\alpha} 5.5^{\alpha - \gamma} (1+z)^{\gamma} \, \, 4.5<z<5
    \end{cases}
\end{equation}
where $\alpha = -0.26$, $\beta = 3.28$, $\gamma = -7.8$, and supernova rate normalization $R_{SN}(0)$ is set to $1.25 \times 10^{-4}\, \text{yr}^{-1}\, \text{Mpc}^{-3}$. 

We explicitly note the presence of a factor of $1/4 \pi$ in Eq.~\eqref{eq-Diffuse-Flux} to account for the differential flux rate per steradian. This factor is crucial to correctly compare the fluence per steradian from Fermi-LAT in Eq.~\eqref{eq:fermiLat} and explains the discrepancy between our work and the previously reported results for diffuse $\gamma$-ray bound in Ref.~\cite{Carenza:2023old}\footnote{After correcting for the larger SN core assumed ($40$ km) in Ref.~\cite{Carenza:2023old}, which gives an additional volume enhancement by a factor of $8$ in production rate.}.  

We can obtain the differential photon flux $\frac{d N_{\gamma}}{d E_{\gamma}}$ from an individual SN by integrating Eq. \eqref{eq:dNdEdt} with respect to time $t_d$,
\begin{equation}
    \begin{split}
    \frac{dN_{\gamma}}{dE_{\gamma}} &= \int dE_{N} d\cos\theta_{N} \frac{1}{2 \gamma (1+ \beta \cos\theta_{N})} \\
    &g_{N}\bigg( \frac{E_{\gamma}}{\gamma (1 + \beta \cos\theta_{N})} \bigg) \exp \bigg(\frac{-R_{env}}{\gamma \beta \tau} \bigg) \frac{dN_{N}}{dE_{N}}.
    \end{split}
\end{equation}
  
Using this expression in Eq.~\eqref{eq-Diffuse-Flux} and requiring it to be below the measured diffuse $\gamma$-ray background in Eq.~\eqref{eq:fermiLat}, we obtain the diffuse $\gamma$-ray constraints shown in Fig. \ref{fig-Diffuse-Constraints} (For discussion on differences with the earlier results reported in literature, see Appendix~\ref{app:comparison}). 

\section{Discussion\label{sec:discussion}}

The primary results of our study are displayed in the exclusion plot for $|U_{\ell 4}|^2$ as function of sterile neutrino mass $M_N$ shown in Fig.~\eqref{fig:U2-Constraints} with solid curves for $|U_{\tau 4}|^2$ and dashed curves for bounds on $|U_{\mu 4}|^2$. The updated constraints from the non-observation of the radiative decay of $N$ from SN1987A by the SMM are shown in blue and the diffuse $\gamma$-ray background constraints from Fermi-LAT are shown in yellow. We also show other existing bounds arising from low-energy SNIIP explosions~\cite{Chauhan:2023sci,Carenza:2023old} (orange), in-flight positron annihilation bound~\cite{DelaTorreLuque:2024zsr} (pink) and bound from BBN assuming standard cosmology~\cite{Boyarsky:2009ix,Ruchayskiy:2012si,Sabti:2020yrt,Domcke:2020ety} (gray). In addition, we also include the parameter region favored by conventional Type-I seesaw models labeled as `SeeSaw' (dark green). The combined existing constraints can reach mixing angles as low as $10^{-13}$. The qualitative and quantitative differences between these constraints and the ones obtained in Ref.~\cite{Carenza:2023old}, are discussed in App.~\ref{app:comparison}. 

The sterile neutrino production rate which is dominated by sterile neutrino upscattering off neutrons, is largely independent of $M_N$ for low $M_N$ but suffers Boltzmann suppression for higher masses. Above the pion mass threshold, the sterile neutrino can efficiently decay into photons through the $N \rightarrow \pi^{0} \nu$. The branching fraction gets enhanced by nearly a factor of $10^3$, which is reflected in the constraint plot by a sudden downward jump of the SMM exclusion curves at $M_N$ just above $135$ MeV. Similar behavior is also seen for the diffuse $\gamma$-ray bound, where the initial flat exclusion region for low masses (up to 50 MeV) becomes weaker due to the Boltzmann suppression in the production rate, leading to an upturn of the diffuse bound for $M_N$ above $50$ MeV. However as soon as $M_N$ becomes greater than $135$ MeV, the pion decay mode leads to a significant increase in the $N$ decay width to photons. This leads to the secluded island in the diffuse $\gamma$-ray constraint shown in Fig.~\ref{fig:U2-Constraints}.   

For very small mixing angles, the decay lengths for most sterile neutrinos that can escape the SN core are larger than the SN envelope. Therefore, the shape of the lower section of the constraint curves are determined by the observation time windows. For observation time of 223s for SMM, the lighter sterile neutrinos decays require higher mixing angles to ensure decay within the observation time. Since heavier sterile neutrinos decay faster, the constraint curves for SMM slope downward. However, since the diffuse $\gamma$-ray constraints integrates the photon flux over an extremely large time window, lighter sterile neutrinos have enough time to decay to contribute to the background photon flux. For the trapping regime in these exclusion curves, the coupling is such that the mean-free-path length becomes less than $R_{\text{env}}$. 

Most importantly, we also display our projections for future exclusion in the $|U_{\ell 4}|^2$-$M_N$ plane, assuming a future galactic supernova at a distance of $10$ kpc from Earth with $\gamma$-ray measurements from Fermi-LAT and e-ASTROGAM. The light green and black curves show the experimental sensitivity of Fermi-LAT for an observation time window of 223s and 3600 s, respectively. For proposed future $\gamma$-ray observatories such as e-ASTROGAM and an observation time window of 3600 s, the exclusion curves are shown in red. A larger effective area and longer observation times can especially help exclude the tiny mixing angles for heavier sterile neutrinos. From Fig.~\ref{fig:U2-Constraints}, we conclude that an observation of a nearby future galactic SN can constrain mixing angles as low as $10^{-16}$ and could provide the leading constraint for $M_N$ up to $700 \text{ MeV}$. 

We also note that the $\gamma$ from lighter sterile neutrinos may lead to the formation of a fireball~\cite{Diamond:2023scc}, which could affect a small portion of the SN1987A constraint from SMM observations. However, we do not perform a detailed study for fireball bound since it can be ruled out from Pioneer Venus Orbiter (PVO) mission~\cite{Diamond:2023scc}.  

\section{Conclusion}
\label{sec:conclusion}
We have examined in detail how $\gamma$-ray data can be used to constrain sterile neutrinos produced in supernovae. First we computed their production rate from a combination of different modes (see Table~\ref{tab:Nprod}), and then found their probability to decay outside the stellar envelope to final states including $\gamma$-ray photons. In particular, we found distinctive features in the time and energy distributions of the decay signal (see e.g. Fig.~\ref{fig-Char-Points}, and Fig~\ref{fig-Char-Points-Extended}). We then proceeded to determine the constraints from existing data, including the SMM observations of SN1987A as well as the diffuse $\gamma$-ray bounds from integrating over all past SNe. To the best of our knowledge, our work is the first to make use of the late-time, post-calibration data from SMM, which can provide complementary constraints to the early-time data (see Fig.~\ref{fig-Early-and-Late}). Then, we examined the bounds from diffuse $\gamma$-ray data and concluded that the bounds exceed those obtained from SMM only at fairly small sterile neutrino masses $\ll 10 $ MeV. Finally, we studied the constraining power of future $\gamma$-ray observations with Fermi-LAT and e-ASTROGAM from a nearby galactic SN (Fig.~\ref{fig:U2-Constraints}), which will exceed the present bounds from SMM by roughly 3 orders of magnitude, offering enormous discovery potential. In the event of a positive detection of an anomalous $\gamma$-ray signal in the next galactic supernova, our results (Figs.~\ref{fig-Char-Points},\ref{fig-Char-Points-Extended}) suggest that the detailed time and energy distribution of the events may contain sufficient information to determine the mass and mixing angle. The ability to reconstruct sterile neutrino properties from a future supernova is probably even more optimistic given that the high-energy neutrino data from Hyper-K may provide additional discovery/constraining potential~\cite{Akita:2023iwq,Telalovic:2024cot}.  We plan to return to these questions in future work. 

For completeness, we also summarize some of the improvements in the calculation of the $\gamma$-ray bound in our work: (1) Rewriting the constraint in Ref.~\cite{Oberauer:1993yr} we find an extra condition on angular variables which affects results at longer times ($\cos{\theta}_{\text{min}}=\text{Max}[-R_{\text{env}}/t_d,-1]$). (see App.~\ref{app:GenExp}) (2) We explicitly include the $\Theta(x-R_{env})$ function to account for the SN envelope size, and demonstrate why using an exponential decay up to $R_{\text{env}}$ to account for modified sterile neutrino flux is not sufficient. The $\Theta$-function explicitly enforces a minimum boost factor (or $E_{N}$) to reach the detector given an observation time ($t_{d}$). (3) Lastly we find that the diffuse flux bound should account for a factor of $1/4\pi$ in the differential photon flux per steradian which was missing in the previous diffuse $\gamma$-ray calculation \cite{Carenza:2023old}. 

\section*{Acknowledgements}
We thank Ying-Ying Li, Vedran Brdar and Leonardo Mastrototaro for helpful discussions. We thank Hans-Thomas Janka and Daniel Kresse for providing the SN profiles used in this work. The work of GC, AG, and IS is supported by the U.S. Department of Energy under the award number DE-SC0020250 and DE-SC0020262. The work of GC is also supported by the NSF Award Number 2309973. AG is partially supported by the World Premier International Research Center Initiative (WPI), MEXT, Japan. AG is grateful to QUP for hospitality during his visit.

\appendix
\section{Simplified General Expression for $\gamma$-Ray Flux}
\label{app:GenExp}
In this section, we present several simplifications of the general expression for the $\gamma$-ray flux. The differential photon flux produced from the decays of the heavy sterile neutrinos, outside the SN envelope is  
\begin{equation}
    \begin{split}
    \frac{d^2 N_{\gamma}}{dE_{\gamma} dt_{d}} &= \int_{-1}^{1} d\cos\theta_{N} \int_{M_N}^{\infty} dE_{N}  \frac{1}{2\tau} g_{N} \bigg(\frac{E_{\gamma}}{\gamma (1 + \beta \cos \theta_{N})} \bigg)\\
    &\times \exp \bigg(\frac{- \gamma t_{d} (1 + \beta \cos \theta_{N})}{\tau} \bigg) \frac{dN_{N}}{dE_{N}} \Theta(x - R_{env}),
    \end{split}
    \label{eq:dNdEdtApp}
\end{equation}
For the general case, this can be simplified further using the $\Theta(x - R_{env})$ condition which leads to
\begin{align}
     \beta t_D  \geq R_{\text{env}} 
     & \implies  \beta t \gamma^2 (1+\beta \cos{\theta}) \geq R_{\text{env}} \nonumber \\
     & \implies \beta^2 (r+c) + \beta -r \geq 0
\end{align}
where $r=R_{\text{env}}/t$ and $c=\cos{\theta}$. The intercepts for the quadratic equation given above are 
\begin{equation}
    \beta_0^{\pm} = \frac{-1 \pm \sqrt{1+4 r (r+c)}}{2(r+c)}
\end{equation}
Let us consider the first case for $(r+c)>0$. Since $c\in [-1,1]$, $\mathcal{D}=1+4 r (r+c)$ cannot be negative. This implies the discriminant is always non-negative and therefore intercepts for the quadratic equation exists. However, we require the solutions to lie between $\beta \in [0,1]$. This implies the condition $c^2+(r+1)c+r\geq 0$ needs to be satisfied. Since the discriminant for this quadratic is also always positive, we need to find the intercepts for $c$, 
\begin{equation}
    c_0^{\pm}=\frac{-(r+1)\pm |r-1|}{2} \implies c_0^{\pm}= -1,-r 
\end{equation}
Therefore for the first case when $(r+c)>0$, we require integration conditions for 
\begin{align}
    c & : [\text{Max}[-r,-1],1] \nonumber\\
    \beta & : [\beta_0^+,1].    
\end{align}
Note that the condition on $\beta$ can be equivalently written as a condition on integration limits for $E_N$. In the second case for $(r+c)<0$, since $c\in [-1,1]$, $\mathcal{D}=1+4 r (r+c)$ cannot be negative, therefore both intercepts exists. It can be shown that $\beta_0^{\pm} \geq 0$, therefore given the same condition on $c$ i.e. $c^2+(r+1)c+r\geq 0$ is satisfied, the required range for the variables $c$ and $\beta$ are 
\begin{align}
    c & : [\text{Max}[-r,-1],1] \nonumber\\
    \beta & : [\beta_0^-,\text{Min}[\beta_0^+,1]].    
\end{align}
but this above derived condition for $c$ conflicts with our original assumption for $(r+c)<0$ and therefore this solution is rejected. 

Therefore, the final simplified expression takes the following form 
\begin{equation}
    \begin{split}
    \frac{d^2 N_{\gamma}}{dE_{\gamma} dt_{d}} &= \int_{c_0}^{1} d\cos\theta_{N} \int_{E_0}^{\infty} dE_{N}  \frac{1}{2\tau} g_{N} \bigg(\frac{E_{\gamma}}{\gamma (1 + \beta \cos \theta_{N})} \bigg)\\
    &\times \exp \bigg(\frac{- \gamma t_{d} (1 + \beta \cos \theta_{N})}{\tau} \bigg) \frac{dN_{N}}{dE_{N}},
    \end{split}
    \label{eq:dNdEdtAppSim}
\end{equation}
where $c_0=\text{Max}[-r,-1]$ and $E_0=\text{Max}[E_N^+,M_N]$
\begin{equation}
    E_N^+=M_N|r+c|\left[\left(\frac{1}{4}+r^2+r\,c\right)^{\frac{1}{2}}-\frac{1}{2}+r\,c+c^2\right]^{-\frac{1}{2}}
    \label{eq:genEmin}
\end{equation}
Note that these integration limits for $\cos{\theta}$ are slightly different than those derived in Ref.~\cite{Oberauer:1993yr}.

\section{Simplified expression for flux from $N\rightarrow \gamma \nu$}
\label{app:Ngammanu}
We can simplify the form of Eq.~\eqref{eq:NnuGMain} further by enforcing the last two $\Theta$-functions, which imposes a lower bound on $E_{N}$ and we obtain
\begin{equation}
    \begin{split}
    \frac{d^2 N_{\gamma}}{dE_{\gamma} dt_{d}} = \int_{E'_{N,\text{min}}}^\infty & dE_{N} \frac{2 E_{\gamma} B_{\gamma \nu}}{\gamma M_N \beta \tau}
    \exp \bigg(\frac{- 2 E_{\gamma} t_{d}}{\tau M_N} \bigg) \frac{dN_{N}}{dE_{N}} \\
    &\times \Theta\left(\frac{2t_d\gamma\beta E_\gamma}{M_N} - R_{env}\right).
    \end{split}
    \label{eq:NnuG2}
\end{equation}
where $E_{N,\text{min}}=\text{Max}[E_{N,\text{min}}^{1},M_N]$ with
\begin{equation}
    E_{N,\text{min}}^1 =\frac{M_N}{2}\frac{E_\gamma^2+E_0^2}{E_\gamma\,E_0}
    \label{eq:Emin1}
\end{equation}
in this case $E_0=M_N/2$. We can simplify Eq.~\eqref{eq:NnuG2} by using the remaining $\Theta$-function which requires
\begin{align}
        \gamma\beta \geq  \frac{ R_{env} M_N}{2 t_d E_\gamma} & \equiv \eta_0 \nonumber\\
        \implies \frac{\beta}{\sqrt{1-\beta^2}} & \geq \eta_0
 \end{align}
We can solve for $\beta$ in the above equation,
\begin{equation}
    \beta^2\geq \frac{\eta_0^2}{1+\eta_0^2}; \implies E_{N,\text{min}}^2 \geq M_N\sqrt{1+\eta_0^2}
\end{equation}
This leads to the final simplified form for the flux
\begin{equation}
    \begin{split}
    \frac{d^2 N_{\gamma}}{dE_{\gamma} dt_{d}} = & \frac{2 E_{\gamma} B_{\gamma \nu}}{ M_N \tau} 
    \exp \bigg(\frac{- 2 E_{\gamma} t_{d}}{\tau M_N} \bigg) \\ & \times \int_{E'_{N,\text{min}}}^\infty dE_{N} \frac{1}{p_N} \frac{dN_{N}}{dE_{N}}
    \end{split}
    \label{eq:NnuG2}
\end{equation}
where $E'_{N,\text{min}}=\text{Max}[E_{N,\text{min}}^{1,2},M_N]$ with
\begin{equation}
    E_{N,\text{min}}^1 =\frac{E_\gamma^2+\left(\frac{M_N}{2}\right)^2}{E_\gamma},\,  E_{N,\text{min}}^2 = M_N\sqrt{1+\left(\frac{ R_{env} M_N}{2 t_d E_\gamma}\right)^2}
    \label{eq:Emin}
\end{equation}

\section{ Results comparison with Ref.~\cite{Carenza:2023old}}
\label{app:comparison}
In this section, we compare our results for SMM and diffuse $\gamma$-ray constraints on the sterile neutrino parameter space with the results reported in the literature recently in Ref.~\cite{Carenza:2023old}. For brevity, we will hereafter refer these previously derived bounds as the \textit{red} curve or with a subscript \textit{old}.
\begin{figure}[ht!]
    \includegraphics[width = \linewidth]{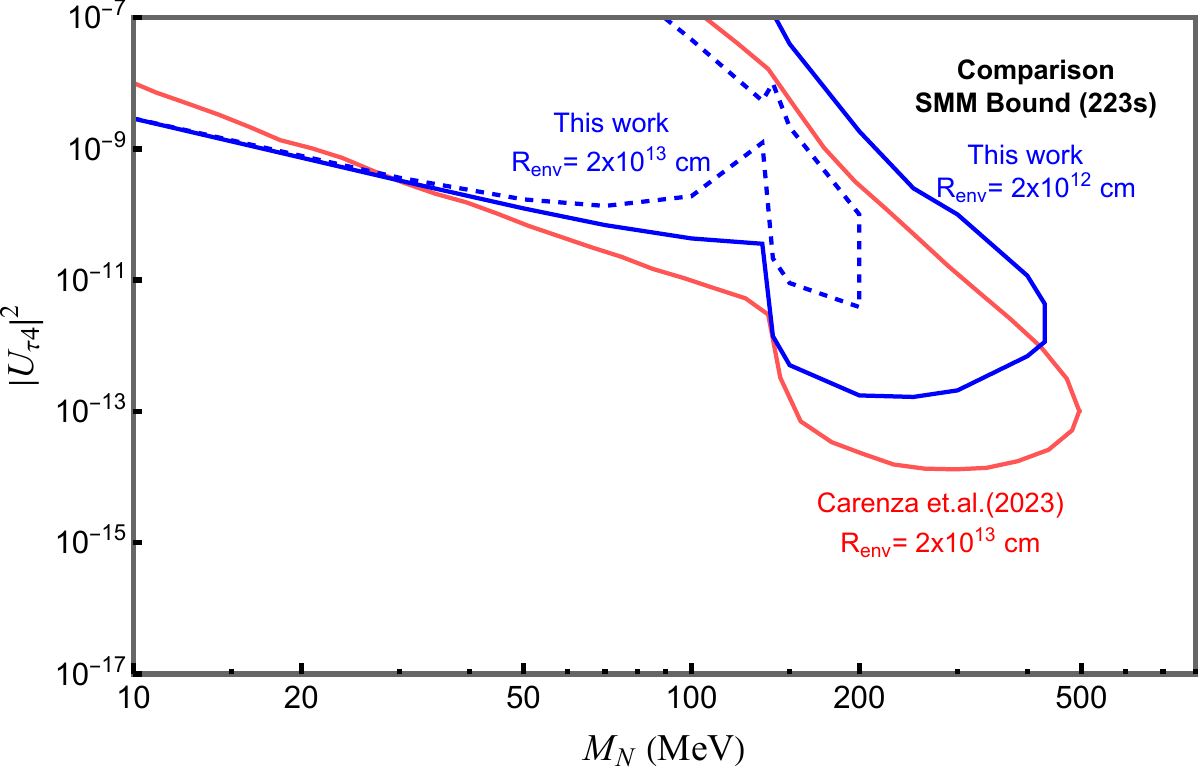}
    \caption{ Comparison of exclusions on $|U_{\tau 4}|^2$ calculated with SMM observations in this work (blue curve) along with result reported in Ref.~\cite{Carenza:2023old} (red curve, for details refer to text). The solid and dashed blue curves correspond to different SN envelope sizes.\label{fig:U2-Constraints-compare}}
\end{figure}
\begin{figure}[ht!]
    \includegraphics[width = \linewidth]{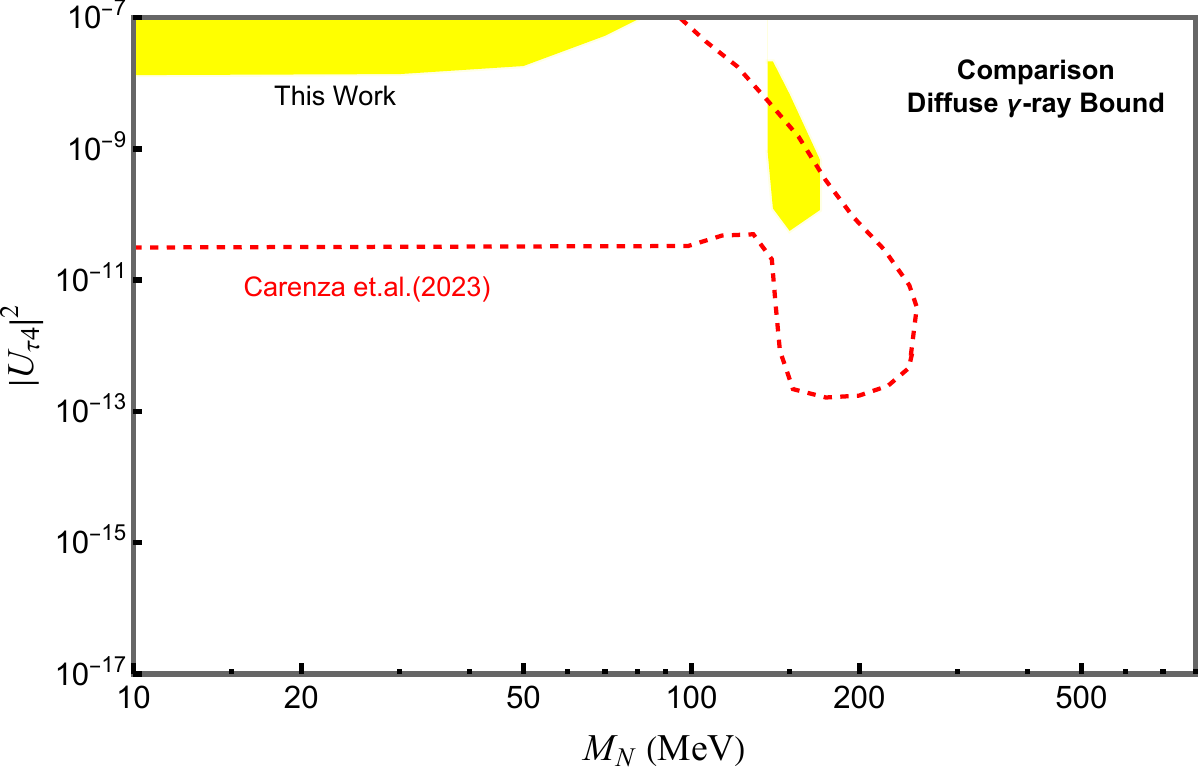}
    \caption{Comparison of diffuse $\gamma$-ray bound on $|U_{\tau 4}|^2$ calculated with Fermi-LAT observations in this work (yellow shaded region) along with result reported in Ref.~\cite{Carenza:2023old} (red dotted curve, for details refer to text).\label{fig-Diffuse-Constraints}}
\end{figure}

The SMM bound comparison is shown in Fig.~\ref{fig:U2-Constraints-compare}. Note we have already discussed the features of the SMM bound shown in Fig.~\ref{fig-Early-and-Late} for different envelope sizes (and observation time) in Sec.~\ref{sec:gammarayobs}. We can clearly see the two major differences between the red and blue curves. 
\begin{itemize}
    \item Firstly, the slope of the exclusion curves for low $M_N$ (below $100$ MeV dominated by $N\rightarrow\gamma\nu$) and low $|U_{\tau4}|^2$ differs. The reason can be attributed to the different dependence of the differential photon flux $\frac{d^2 N_{\gamma}}{dE_{\gamma} dt_{d}}$ on $M_N$. 
    \begin{equation}
    \begin{split}
    \left(\frac{d^2 N_{\gamma}}{dE_{\gamma} dt_{d}}\right)_{\text{Old}} &=  \frac{ M_N B_{\gamma \nu}}{ 2 E_{\gamma} \tau} 
    \exp \bigg(\frac{- 2 M_N t_{d}}{\tau E_{\gamma}} \bigg) \\ & \times \int_{\text{Max}[E_{N,\text{min}}^{1},M_N]} dE_{N} \frac{1}{p_N} \frac{dN_{N}}{dE_{N}}
    \end{split}
    \label{eq:NnuGCarenza}
    \end{equation}
    where $E_{N,\text{min}}^1 =\frac{E_\gamma^2+\left(\frac{M_N}{2}\right)^2}{E_\gamma}$. Note that the flux in this expression is non-zero at $t=0$ for even heavier $M_N$.

    For completeness, we restate the correct expression used in this work (see App.~\ref{app:Ngammanu} for derivation)
    \begin{equation}
    \begin{split}
    \frac{d^2 N_{\gamma}}{dE_{\gamma} dt_{d}} &=  \frac{2 E_{\gamma} B_{\gamma \nu}}{ M_N \tau} 
    \exp \bigg(\frac{- 2 E_{\gamma} t_{d}}{\tau M_N} \bigg) \\ & \times \int_{E'_{N,\text{min}}}^\infty dE_{N} \frac{1}{p_N} \frac{dN_{N}}{dE_{N}}
    \end{split}
    \label{eq:NnuG2}
\end{equation}
where $E'_{N,\text{min}}=\text{Max}[E_{N,\text{min}}^{1,2},M_N]$ where $ E_{N,\text{min}}^2 = M_N\sqrt{1+\left(\frac{ R_{env} M_N}{2 t_d E_\gamma}\right)^2}$.

    \item Secondly, the reach for the $|U_{\tau4}|^2$ at higher $M_N$ is better for the red curve. This can attributed to the following reason : The red curve does not explicitly uses a $\Theta(x-R_{env})$ function to account for time of delay of particles propagating inside the envelope. Instead, they choose to only take into account the fraction of sterile neutrinos decaying outside the envelope,
    $$\left(\frac{dN_4}{dE_N}\right)_{esc}= \frac{dN_4}{dE_N} \text{exp}\left(-\frac{R_{env}}{\lambda_{\text{SN}}}\right)$$
    where $\lambda_{\text{SN}}=\gamma\beta\tau$. But the above expression fails to account for the corresponding effect on $t_d$, which is quite important for heavier $M_N$ and shorter observation times.  Upon proper accounting for this delay, given some envelope size and a specified $t_d$, only a certain minimum energy particles can reach the detectors on earth. For example,  this corresponds to the condition $E_{N,\text{min}}^2$ in Eq.~\eqref{eq:Emin} for $N\rightarrow\gamma\nu$ and more generally specified by Eq.~\eqref{eq:genEmin}. We also note that the SMM observations lasted for 223.2 seconds, while the red curve uses an observation time of 232.2 seconds.
\end{itemize}

The diffuse $\gamma$-ray bound comparison is shown in Fig.~\ref{fig-Diffuse-Constraints}. We have already discussed the features of this  bound derived in our work, in Sec.~\ref{sec:diffusegammaray}. While the shape of the exclusion areas are quite similar i.e. flat exclusion region for low $M_N$ and a sudden increase in the sensitivity after the pion mass threshold ($M_N > 135$ MeV), we can clearly see a major difference between the quantitative nature of the yellow and red curves. We find constraints from diffuse $\gamma$-rays to be much weaker than those published in Section III E of \cite{Carenza:2023old}. This is attributed to two distinct reasons : 
\begin{itemize}
    \item We note that our Eq. \ref{eq-Diffuse-Flux} includes a factor of $\frac{1}{4 \pi}$ missing in Eq.(3.27) of \cite{Carenza:2023old}. This is an important factor to account for the differential flux per steradian before comparing to the measurements from Fermi-LAT. 
    \item The sterile neutrino production rate is directly proportional to the SN core volume is  i.e. $dN/dE\propto R_{\text{SN}}^3$. In our work, we have used typical $R_{\text{SN}} \sim 20$ km, while the red curve uses a $R_{\text{SN}} \sim 40$ km. Since the bounds are directly proportional to the total sterile neutrino production rate, there is a factor of $8$ difference from the volume enhancement in the exclusion curves. 
\end{itemize}
Therefore upon combining the two enhancements, we expect the red curve to be naively stronger than the yellow curve by a factor of $32\pi\sim 100 $ or more.

We observe that the diffuse $\gamma$-ray bounds in our work is far weaker than our SMM bound. We now argue that this is the expected behavior in a general case, with an analogy with SN neutrino flux. 
It is well known that a single nearby galactic supernova such as SN1987A should have a direct neutrino flux many orders of magnitude greater than the expected diffuse flux from DSNB~\cite{Lunardini:2009ya,Beacom:2010kk}. This is also evident from the observation of neutrino flux from SN1987A, whereas the DSNB still remains elusive. Therefore, the only way a diffuse $\gamma$-ray background from sterile neutrino decays could overwhelm the direct expected flux in SMM during SN1987A is if the sterile neutrino is so long-lived that only an incredibly small fraction of the sterile neutrinos produced decay within the 223.2s SMM observation time window. Since this is in conflict with our obtained bounds from SMM, the diffuse $\gamma$-ray flux bound should be comparatively much weaker, as shown in our work.

\section{Effect of limited angular resolution}
The future observations of galactic SN may run for longer times, and thus we might have to worry about an angular spread in the $\gamma$-rays. Consider a sterile neutrino emitted from a SN at a distance $D_{SN}$ away from Earth. The sterile neutrino travels a distance $x$ before decaying, and the observation angle is taken to be $\phi$. We can show that that the time delay relative to the neutrino observation is
\begin{equation}
    t_{d} = \frac{x}{\beta} - \sqrt{x^2 - D_{SN}^2 \sin^2\phi} - D_{SN} (1 - \cos\phi).
\end{equation}
Inverting this equation to solve for the observation angle in terms of the time of delay, we find
\begin{equation}
    \cos\phi = \frac{\beta^2 (2 D_{SN}^2 + 2 D_{SN} t_{d} + t_{d}^2 - x^2) - \beta x (D_{SN} + t_{d}) + x^2}{2 \beta D_{SN}(-x+\beta(D_{SN} + t_{d}))}.
\end{equation}
Given the limited observation time window for any given $\gamma$-ray observatory, and assuming that the sterile neutrinos are not long-lived on galactic time scales (i.e. $d \gg t_{d}, x$). In this limit
\begin{equation}
    \cos\phi \approx 1 + \frac{\beta^2 t_{d}^2 - 2 \beta t_{d} x - \beta^2 x^2 + x^2}{2 \beta^2 d^2}
\end{equation}
We can see that the above expression reaches an extremum when $x = t_{d} \gamma^2 \beta$. Since for small angles $\cos\phi \approx 1 - \frac{1}{2} \phi^2$, we can set a limit on the observation angle of
\begin{equation}
    \phi \lsim \frac{\gamma \beta t_{d}}{d} \leq \frac{M_N t_{d}}{E_{N, \max} d}.
\end{equation}
since $t_{d} \ll d$, this will result in very little angular spread in the observations.

\section{Effect of progenitor opacity}
In this work, we have made the assumption that $\gamma$-rays produced inside of the envelope are fully attenuated, while those produced outside experience no attenuation. Due to the opaqueness of the progenitor envelope, there will be an additional effect: $\gamma$-rays which propagate backwards relative to the parent sterile neutrino emitted away from the Earth, might enter the envelope and get absorbed.

To put it in terms of $x$ and $\theta_{s}$ as defined in Sec. \ref{sec:Geom}, the closest our $\gamma$-ray gets to the SN center is
\begin{equation}
    \begin{cases}
        r_{\min} = x \, \mathrm{for} \, \theta_{s} < \frac{\pi}{2} \\
        r_{min} = x \sin(\pi - \theta_{s}) \, \mathrm{for} \, \theta_{s} > \frac{\pi}{2}
    \end{cases}
\end{equation}
and we can account for this by replacing the $\Theta(x-R_{env})$ in Eq. \ref{eq:dNdEdt} with $\Theta(r_{\min} - R_{env})$. We have checked that this replacement does not noticeably change our results.

\section{Production rate matrix elements and decay widths}
\label{app:decaywidths}
The sterile neutrino production rates depend on the corresponding matrix elements for each process given in Tab.~
\ref{tab:Nprod}. We provide the matrix elements in in Tab. \ref{table:matrixElements}, taken from Ref.~\cite{Carenza:2023old}. Note the matrix coefficients used have the following values,
\begin{equation}
    g_L=-0.5+\sin^2{\theta_W},\,g_R=\sin^2{\theta_W}
\end{equation}
and for lepton scattering with nucleon,  
\begin{equation}
    G_1=2(G_A+G_V)^2,\, G_2=2(G_A-G_V)^2,\,G_3=2(G_A^2-G_V^2)
\end{equation}
where for the neutral current interaction, 
\begin{eqnarray}
   G_V&\to& G_V^n=\frac{1}{2} \,\,\,\,\,\, , \,\,\, G_V^p=\frac{1}{2}-2\sin^2\theta_W \,\ , \\
    G_A&\to& G_A^n=\frac{g_A}{2} \,\ , \,\ G_A^p=\frac{g_A}{2} \,\ .
\end{eqnarray}
and for the charged current process, we use 
\begin{eqnarray}
     G_V\to V_{ud}G_V,\quad G_V&=&\frac{g_V\left(1-\frac{q^2(\gamma_p-\gamma_n)}{4M_n^2}\right)}{\left(1-\frac{q^2}{M_n^2}\right)\left(1-\frac{q^2}{M_V^2}\right)^2} \,\ , \\
    G_A\to V_{ud}G_A,\quad G_A&=&V_{ud}\frac{g_A}{
    \left(1-\frac{q^2}{M_A^2}\right)^2} \,\ . 
\end{eqnarray}
here the variables $G_{V,A}$ with superscript $(n,p)$ stands for vector and axial charges for neutron and proton respectively, $g_V=1$ and $g_A=1.27$ are the vector and axial vector coupling constant, respectively. Moreover, $\gamma_{p/n}$ are the proton(neutron) magnetic moment, $M_V=840$ MeV is the vector mass, $M_A=1$ GeV is the axial mass and $M_n$ is the effective nucleon mass inside the SN core. 

\begin{table*}[!h]
\centering
\begin{tabular}{|c | c|} 
 \hline
 Process ($1+2\rightarrow 3+4$) & $|M|^2$/($8 G_F^2 |U_{\ell 4}|^2$) \\ [0.5ex] 
 \hline\hline
 $\nu_{\ell}+\bar{\nu}_{\ell}\leftrightarrow\bar{\nu}_{\ell}+N$  &$8(p_1\cdot p_3)(p_2\cdot p_4)$ \\
 \hline
$\nu_{\ell}+\nu_{\ell}\leftrightarrow\nu_{\ell}+N$ & $4(p_1\cdot p_2)(p_3\cdot p_4)$\\
\hline
$\nu_{\ell'}+\bar{\nu}_{\ell'}\leftrightarrow\bar{\nu}_{\ell}+N$ & $2(p_1\cdot p_3)(p_2\cdot p_4)$\\
\hline
$\nu_{\ell}+\bar{\nu}_{\ell'}\leftrightarrow\bar{\nu}_{\ell'}+N$  & $2(p_1\cdot p_3)(p_2\cdot p_4)$\\
\hline
$\nu_{\ell}+\nu_{\ell'}\leftrightarrow\nu_{\ell'}+N$ & $2(p_1\cdot p_2)(p_3\cdot p_4)$\\
\hline
$\nu_\ell+n\leftrightarrow n+N$ &  \\
$\nu_\ell+p\leftrightarrow p+N$ &  $ G_1(p_1\cdot p_2)(p_3\cdot p_4) + G_2(p_2\cdot p_4)(p_1\cdot p_3) - G_3\,m_2\,m_3\,(p_1\cdot p_4)$ \\
$\mu^-+p\leftrightarrow n+N$ &\\
\hline
$\mu^-+\nu_e\leftrightarrow e^-+N$ & $8(p_1\cdot p_2)(p_3\cdot p_4)$ \\
 \hline
 $\nu_{\ell}+e^-\leftrightarrow e^-+N$ & $8 [g_L^2(p_1\cdot p_2)(p_3\cdot p_4) + g_R^2(p_1\cdot p_3)(p_2\cdot p_4)-g_Lg_Rm_e^2(p_1\cdot p_4)]$  \\ 
 \hline
$\nu_{\ell}+e^+\leftrightarrow e^++N$ & $8 [g_R^2(p_1\cdot p_2)(p_3\cdot p_4) + g_L^2(p_1\cdot p_3)(p_2\cdot p_4)-g_Lg_Rm_e^2(p_1\cdot p_4)]$  \\ 
 \hline
  $e^++e^-\leftrightarrow\bar{\nu}_{\ell}+N$ & $8 [g_R^2(p_1\cdot p_4)(p_2\cdot p_3) + g_L^2(p_1\cdot p_3)(p_2\cdot p_4)-g_Lg_Rm_e^2(p_3\cdot p_4)]$  \\ 
 \hline
\end{tabular}
\caption{Matrix element squared $|M|^2$ for the processes involved in sterile neutrino production in units of $8 G_F^2 \sin^2\theta_\tau$~\cite{Fuller:2008erj,Carenza:2023old}. All matrix elements coefficients $(g_{L,R},G_{1,2,3})$ are provided in App.~\ref{app:decaywidths}.\label{table:matrixElements}}

\end{table*}

Our expression for the decay widths is given in Table \ref{tab:Decay_Rates}, taken from Ref.~\cite{Coloma:2020lgy}. In this, we treat electrons as massless. Since we are considering Majorana sterile neutrinos, the charge conjugated version of all listed process are also available, so the total decay rate is found by multiplying by 2.
\begin{table*}[ht]
    \centering
    $\begin{array}{| c|c|c|}
    \hline 
    \text{Process} & \Gamma/(G_F^2 |U_{\ell 4}|^2 M_{N}^{3}) &\text{Requirements} \\ \hline
    N \rightarrow  \nu_{\alpha} \gamma & 9 \alpha_{EM} M_N^2/(2048 \pi^4)& \text{None} \\
    N \rightarrow \nu_{\alpha} \nu_{\beta} \bar{\nu_{\beta}} & M_{N}^2 /(192 \pi^3) & \text{None} \\
    N \rightarrow \nu_{\alpha} e^{-} e^{+} & (\frac{1}{4} - \sin^2\theta_{W}+2\sin^{4} \theta_{W}) M_{N}^2/(192 \pi^3) & \text{None} \\
    N \rightarrow \nu_{e} e^{+} \mu^{-} & \frac{M_{N}^2}{384 \pi^3} \bigg( 2 (1 - x_{\mu}^2)(2 + 9 x_{\mu}^2) + 2 x_{\mu}^2 (1-x_{\mu}^2)(-6 -6 x_{\mu}^2 +x_{\mu}^4 + 6 \ln{x_{\mu}^2}) \bigg) & \alpha = \mu, M_{N} > m_{\mu} \\
    N \rightarrow \nu_{\alpha} \pi^{0} & f_{\pi}^2 *(1 - x_{pi}^2)^2/(32 \pi) & M_{N} > m_{\pi} \\
    N \rightarrow \mu^{-} \pi^{+} & \frac{f_{\pi}^2 |V_{ud}|^2}{16 \pi} \lambda^{1/2}(1, x_{\pi^{+}}^2, x_{\mu}^2) \big(1 - x_{\pi^{+}}^2 - x_{\mu}^2 (2 + x_{\pi^{+}}^2 - x_{\mu}^2) \big) & \alpha = \mu, M_{N} > m_{\mu} + m_{\pi^{+}} \\
    \hline
    \end{array}$
    \caption{Decay rates for the various processes considered in this work, obtained from \cite{Coloma:2020lgy}. The expression $x_{i} = m_{i}/M_N$, while $\lambda(a,b,c)$ is the Kallen function. We take constants $\alpha_{EM} = 1/137$, $\sin^2 \theta_{W} = 0.23$, $f_{\pi} = 135$ MeV, and $V_{ud} = 0.974$. \label{tab:Decay_Rates}}
\end{table*}

\bibliographystyle{apsrev4-1.bst}
\bibliography{bibliography.bib}

\end{document}